\documentclass[final]{aipproc}

\usepackage{amssymb}

\layoutstyle{6x9}

\def\Dslash{\hspace{3pt}\raisebox{1pt}{$\slash$} \hspace{-7.5pt} D}

\begin{document}

\title{Gauge-Higgs Unification on Flat Space Revised}

\classification{11.10.Kk, 11.15.Ex, 12.60.-i}
\keywords      {Field Theories in Higher Dimensions, Beyond the Standard Model, Spontaneous Electroweak Symmetry Breaking}

\author{Giuliano Panico}{
  address={ISAS-SISSA and INFN, Via Beirut 2-4, I-34013 Trieste, Italy}
}

\begin{abstract}
 Models with gauge-Higgs unification on a flat space are typically affected by
common problems, the main of which are the prediction of a too small top and
Higgs mass and a too low compactification scale. We show how, by breaking the
SO(4,1) Lorentz symmetry in the bulk and introducing a ${\bf Z}_2$ ``mirror'' symmetry,
a potentially realistic model arises, in which all these problems are solved.
\end{abstract}

\maketitle

%%%%%%%%%%%%%%%%%%%%%%%%%%%%%%%%%%%%%%%%%%%%
%% MAINMATTER
%%%%%%%%%%%%%%%%%%%%%%%%%%%%%%%%%%%%%%%%%%%%

\section{Introduction}

The idea of identifying the Higgs field with the internal components of a gauge field in
TeV--sized extra dimensions (Gauge-Higgs Unification (GHU)) has been realized to offer
a possible solution to the SM instability of the electroweak scale. However GHU models
compactified on a flat $S^1/{\bf Z}_2$ orbifold usually present some common drawbacks: too small
Higgs and top masses and a too low compactification scale \cite{Scrucca:2003ra}.

To solve these problems we propose to break the Lorentz symmetry in the bulk along
the fifth direction
\cite{Panico:2005dh}.\footnote{In the following we consider an explicit breaking of
the Lorentz symmetry. However, in a purely $5D$ context, it can be generated by
an axion--like field with twisted boundary conditions, as shown in \cite{Panico:2005dh}.}
In this way, Yukawa couplings are no longer constrained by the
$5D$ Lorentz invariance and one can obtain order one couplings which are needed to get a
realistic top mass. This leads also to an enhancement of the Higgs quartic coupling
which pushes the Higgs mass above the current experimental bounds ($m_H > 115$ GeV).

Moreover, the introduction of a ${\bf Z}_2$ ``mirror'' symmetry, which doubles a subset
of the bulk degrees of freedom, allows for a partial cancellation of the mass term in
the effective Higgs potential, thus generating a substantial gap between the 
electroweak and compactification scales.

\section{The model}

We consider a 5D gauge theory compactified on $S^1/{\bf Z}_2$. The gauge group is
taken to be $SU(3)_w\times G_1\times G_2$, where $G_i = U(1)_i \times SU(3)_{i,s}$, $i=1,2$,
with the requirement of having a Lagrangian invariant under the ${\bf Z}_2$ ``mirror'' symmetry
$1\leftrightarrow 2$. The ${\bf Z}_2$ orbifold projection is embedded non-trivially in the
$SU(3)_w$ group, by means of the diagonal matrix $P=diag(-1,-1,1)$; this choice of parities
leads to the breaking $SU(3)_w \rightarrow SU(2)_L \times U(1)_w$. For the Abelian $U(1)_i$
and non-Abelian $SU(3)_{1,s}$ fields, we take
$A_1(y\pm 2 \pi R) = A_2(y)$ and $A_1(-y)=\eta A_2(y),$
where $\eta_\mu = 1$, $\eta_5=-1$. The unbroken
gauge group at $y=0$ is $SU(2)\times U(1)\times G_+$, whereas at $y=\pi R$ we have
$SU(2)\times U(1)\times G_1 \times G_2$, where $G_+$ is the diagonal subgroup of $G_1$ and $G_2$.
The ${\bf Z}_2$ mirror symmetry also survives the compactification and remains as an exact
symmetry. The boundary conditions
can be diagonalized by
a redefinition of the bulk fields $A_\pm = (A_1\pm A_2)/\sqrt{2}$; these combination are
respectively periodic and antiperiodic on $S^1$ and have a multiplicative charge $+1$ for $A_+$
and $-1$ for $A_-$ under the mirror symmetry.
The 4D zero-modes
are the gauge fields in the adjoint of $SU(2)\times U(1) \in SU(3)_w$, the $U(1)_+$ and gluon
gauge bosons $A_+$ and a charged scalar doublet identified with the Higgs field, which arises
from the even internal components of the $SU(3)_w$ 5D gauge fields (namely $A_w^{4,5,6,7}$).
The $SU(3)_{+,s}$ and $SU(2)$ gauge groups are identified respectively with the SM $SU(3)_{QCD}$
and $SU(2)_L$ ones, while the hypercharge $U(1)_Y$ is the diagonal subgroup of $U(1)_w$ and
$U(1)_+$. The extra $U(1)_X$ gauge symmetry which survives the orbifold projection is anomalous
(see \cite{Scrucca:2003ra, Panico:2005dh, Panico:2006em}) and its corresponding gauge boson
decouples from the low-energy effective theory. This mechanism is required in order to obtain
the correct value of the weak mixing angle. The most general gauge Lagrangian compatible with
the mirror symmetry and the 4D Lorentz symmetry is
\begin{equation}
{\cal L}_g = \sum_{i=1,2} \left[-\frac{1}{4} F_{i\mu\nu} F^{i\mu\nu}
- \frac{\rho_s^2}{2} F_{i\mu 5} F^{i\mu 5}\right]
-\frac{1}{2} {\rm Tr}\, F_{\mu\nu} F^{\mu\nu} - \rho_w^2 {\rm Tr}\, F_{\mu 5} F^{\mu 5},
\end{equation}
where we omitted for simplicity the gluon fields, the ghost Lagrangian and the gauge-fixing
terms.

An additional spontaneous symmetry breaking to $U(1)_{EM}$ is induced by an Higgs VEV
$\langle A_{w y} \rangle = \frac{2 \alpha}{g_5 R} t_7$, where $g_5$ is the 5D
charge of the $SU(3)_w$ group and $t_a$ its generators.
The Higgs VEV is associated to a Wilson line $W=e^{4 \pi i \alpha t_7}$, thus the electroweak
symmetry breaking (EWSB) is equivalent to a Wilson line symmetry breaking \cite{hos}.

In the bulk we introduce couples of fermions $(\Psi, \widetilde \Psi)$, with
identical quantum numbers and opposite orbifold parities. There are couples
$(\Psi_1, \widetilde \Psi_1)$ which are charged under $G_1$ and neutral under $G_2$ and,
by mirror symmetry, an equal number of couples $(\Psi_2, \widetilde \Psi_2)$ charged under $G_2$
and neutral under $G_1$. For each quark family we include one pair of couples
$(\Psi_{1,2}^t, \widetilde \Psi_{1,2}^t)$ in the $\overline{\bf 3}$ representation of $SU(3)_w$
and one pair of couples $(\Psi_{1,2}^b, \widetilde \Psi_{1,2}^b)$ in the ${\bf 6}$
representation of $SU(3)_w$. Both pairs have $U(1)_{1,2}$ charge $+1/3$ and are in the ${\bf 3}$
representation of $SU(3)_{1,2,s}$. These fields satisfy twisted boundary condition similar to
the gauge field ones,
in particular the combinations
$\Psi_\pm = (\Psi_1 \pm \Psi_2)/\sqrt{2}$ are respectively periodic and antiperiodic on $S^1$.
The bulk fermion Lagrangian for a quark family has the form
\begin{equation}
{\cal L}_\Psi = \sum_{{i=1,2}\atop{a=t,b}} \left\{ \overline{\Psi}_i^a
\left[i \Dslash_4 - k_a D_5 \gamma^5\right]\Psi_i^a
+ \overline{\widetilde\Psi}_i^a \left[i \Dslash_4 - \widetilde{k}_a D_5 \gamma^5\right]
\widetilde\Psi_i^a
- M_a \left(\overline{\widetilde\Psi}_i^a \Psi_i^a + {\overline\Psi}_i^a
\widetilde\Psi_i^a\right)\right\}.
\end{equation}

We also introduce massless fermions localized at the $y=0$ fixed point with mirror charge $+1$.
These fields can couple only with periodic even bulk fields through localized mass terms (see
\cite{Panico:2005dh, Panico:2006em} for a detailed description of the couplings). In particular
we put an $SU(2)_L$ doublet $Q_L$ and two singlets $t_R$ and $b_R$, all in the fundamental
representation of $SU(3)_{+,s}$ and with $U(1)_+$ charge $+1/3$.

\section{The Electroweak Symmetry Breaking}

The 5D gauge symmetry, which remains unbroken, forbids the appearance of any local Higgs
potential in the bulk, moreover a localized Higgs potential is also forbidden.
The Higgs potential is
radiatively generated and induced by non-local operators, thus it is finite. The contributions
of the various fermionic and gauge fields to the effective potential are fully reported in
\cite{Panico:2005dh, Panico:2006em}.
The whole effective potential is dominated by the contributions of the top and bottom
quark, which provide a negative contribution for the Higgs mass squared needed to
generate a spontaneous EWSB.
The presence of an unbroken mirror symmetry forces bulk fermions to come in pairs of periodic
and antiperiodic fields with the same quantum numbers and the same Lorentz-breaking couplings.
This allows for a natural cancellation of the radiatively induced Higgs mass, lowering the
position of the minimum of the effective potential and reducing the amount of fine-tuning needed
to satisfy the experimental bounds \cite{Panico:2006em}.

After EWSB the W boson zero-mode acquires a mass
$m_W = \alpha/R$. The whole spectrum of the fermion system is described in 
\cite{Panico:2005dh, Panico:2006em}. In particular the top mass, in the limit of
large bulk-boundary mixing
and small $\alpha$, is given by
\begin{equation}
m_t \simeq k_t m_W \frac{2 \lambda_t/k_t}{\sinh(2 \lambda_t/k_t)},
\end{equation}
where $\lambda_t = \pi R M_t$. From this relation we deduce that $k_t \sim 2 - 3$ is
needed to get the correct top mass.
The bulk-boundary fermion system provides also the lightest non-SM states. These states
are coloured fermions with a mass of order $1-2$ TeV, and in particular they are
given by an $SU(2)$ triplet (with $Y=2/3$), a doublet (with $Y=-1/6$) and a singlet
(with $Y=-1/3$).

\section{Phenomenological Bounds}

Flavour and CP conserving new physics effects can be parameterized by the universal parameters
$\widehat S$, $\widehat T$, $\widehat U$, $V$, $X$, $W$ and $Y$, and by 3 parameters which
describe the distortions of the $\widehat Z$ boson couplings with the quarks
\cite{Cacciapaglia:2006pk}.

The light quarks are nearly exactly localized, thus their couplings with the $\widehat Z$ are
undistorted. The third quark family has a stronger mixing with the bulk fields.
Available experimental data put tight bounds on the $\widehat Z \bar b_L b_L$ vertex,
which, in our model, gets corrections due to two effects.
One effect is the
mixing of $b_L$ with the KK tower of an $SU(2)_w$ triplet and an $SU(2)_w$ singlet bulk fermions.
However this distortion is much suppressed due to the choice of the bulk
fermion quantum numbers and vanishes in the limit of zero $b$ mass \cite{Panico:2006em}.
The other distortion
comes from the couplings of the $b_L$ field with the KK tower of the anomalous $U(1)_X$
gauge subgroup. The correction turns out to be $-\delta g_b \lesssim 3 \alpha^2$, and
can be neglected for $\alpha \lesssim 3 \times 10^{-2}$.

Out of the 7 universal parameters, only $\widehat S$, $\widehat T$, $Y$ and $W$ arise
at leading order in $\alpha$:
\begin{equation}
\left\{
\begin{array}{l}
\widehat S = \frac{2}{3} \alpha^2 \pi^2\\
\rule{0pt}{1.25em}\widehat T = \alpha^2 \pi^2
\end{array}
\right.\,,
\qquad \qquad
\left\{
\begin{array}{l}
Y \simeq \frac{1}{3} \alpha^2 \pi^2\\
\rule{0pt}{1.25em}W = \frac{1}{3} \alpha^2 \pi^2
\end{array}
\right.\,,\label{eqUnivPar}
\end{equation}
the others are suppressed by higher powers of $\alpha$ and can be neglected in the fits.

\begin{figure}
  \includegraphics[height=.3\textheight]{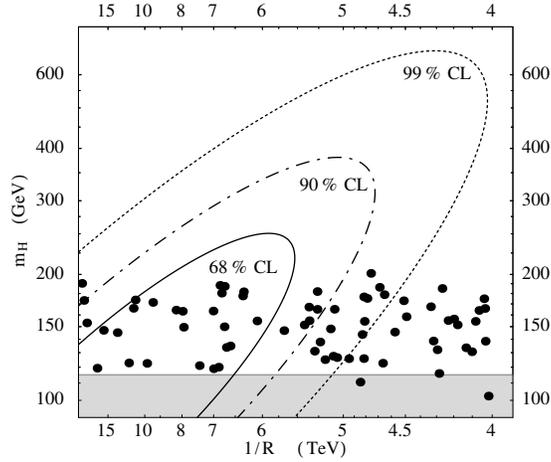}
  \caption{Constraints coming from a $\chi^2$ fit on the 
electroweak precision tests. The shaded band
shows the experimentally excluded values for the Higgs mass ($m_H<115$ GeV).
The dots represent
the predictions of our model for different values of the microscopic parameters.}
\label{figEWPTfit}
\end{figure}

In fig.~\ref{figEWPTfit}, we report the constraints on the Higgs mass and the
compactification scale obtained by a $\chi^2$ fit using the values in eq.~(\ref{eqUnivPar}).
From the fit, one can extract a lower bound on the compactification scale
$1/R \gtrsim 4 - 5$ TeV (which corresponds to $\alpha \lesssim 0.016 - 0.02$) and an upper bound
on the Higgs mass which varies from $m_H \lesssim 600$ GeV at $1/R \sim 4$ TeV to
$m_H \lesssim 250$ GeV for $1/R \gtrsim 10$ TeV.
Notice that the values for $\widehat S$, $W$ and $Y$ found in our model
are essentially those expected in a generic $5D$ theory with gauge bosons and Higgs
in the bulk \cite{Barbieri:2004qk}. On the contrary, the $\widehat T$ parameter
is much bigger and is essentially due to the anomalous gauge field $A_X$, which
generates a distortion of the SM $\rho$ parameter. The high value of the
$\widehat T$ parameter compensates the effects of a heavy Higgs.

Performing a random scan of the microscopic parameters (see fig.~\ref{figEWPTfit}),
we found that the bound on
the compactification scale can be easily fullfilled if a certain amount of
fine-tuning is allowed. We estimated the fine-tuning to be of order a few \%
\cite{Panico:2006em}.

\bibliographystyle{aipproc}

\end{document}